\documentclass[10pt, journal, draftclsnofoot, onecolumn]{IEEEtran}
\usepackage{citesort}
\usepackage{color}
\usepackage[latin9]{inputenc}
\usepackage{amsthm}
\usepackage{amsmath}
\usepackage{amsthm}
\usepackage{multirow}
\usepackage{amssymb}
\usepackage{graphicx}
\usepackage{lipsum}
\usepackage{algorithmic}
\usepackage{algorithm}
\usepackage{mathrsfs}

\theoremstyle{definition}

\theoremstyle{remark}

\begin{document}

\title{Extracting and Exploiting Inherent Sparsity for Efficient IoT Support in 5G: Challenges and Potential Solutions}
\author{Bassem~Khalfi,~\IEEEmembership{Student~Member,~IEEE,}
        ~Bechir~Hamdaoui,~\IEEEmembership{Senior~Member,~IEEE,}
        and~Mohsen~Guizani,~\IEEEmembership{Fellow~Member,~IEEE}}

\maketitle

\begin{abstract}
Besides enabling an enhanced mobile broadband, next generation of mobile networks (5G) are envisioned for the support of massive connectivity of heterogeneous Internet of Things (IoT)s. These IoTs are envisioned for a large number of use-cases including smart cities, environment monitoring, smart vehicles, etc. Unfortunately, most IoTs have very limited computing and storage capabilities and need cloud services. Hence, connecting these devices through 5G systems requires huge spectrum resources in addition to handling the massive connectivity and improved security.
This article discusses the challenges facing the support of IoTs through 5G systems.
The focus is devoted to discussing physical layer limitations in terms of spectrum resources and radio access channel connectivity. We show how sparsity can be exploited for addressing these challenges especially in terms of enabling wideband spectrum management and handling the connectivity by exploiting device-to-device communications and edge-cloud. Moreover, we identify major open problems and research directions that need to be explored towards enabling the support of massive heterogeneous IoTs through 5G systems.
\end{abstract}

\section{Introduction}
\label{sec:introduction}
The foreseen success of the Internet of Things (IoT) and its applications is the result of three major trends.
First, fifth-generation (5G) systems have promises for meeting stringent QoS requirements that legacy systems fail to meet. Examples of such requirements are high data rates, low energy consumption, low latency, high capacity, and improved security. These expected improvements make 5G an ideal candidate for ensuring required connectivity and services for massive and heterogenous IoT devices~\cite{3GPP-2016}.
The second trend is the emergence of cloud computing services, which are believed to play a vital role in making IoT a success by enabling diverse IoT services and applications that were not possible before. Bringing computing and storage resources closer to the IoT devices by means of edge computing has great promises for lowering energy consumption by releasing the devices from the burden of having to deal with some or most of the  computation and energy needed for task execution~\cite{bonomi2012fog}.
The third trend is the adoption of device-to-device (D2D) communications 
envisioned for public safety with the potential for enabling more decentralized network management and local traffic offloading~\cite{dohler20165g}. D2D offers 5G real-time assurances and better spectrum and resource allocation efficiency~\cite{dohler20165g}.
These (technological) trends have jointly led to a common belief that the success of IoT applications is rather a possible reality.

Indeed, telecom industries believe that IoT is the main driver of 5G, as the major use-cases for 5G involve IoT devices (e.g., consumer or industrial IoTs). For instance, IoT will shift the focus of mobile system designs from enabling traditional broadband communications to support not only enhanced broadband communications but especially massive IoTs with heterogeneous services requirements.
However, there are major challenges that need to be addressed in order for 5G to support these massive IoTs. The first challenge lies in the enormous amounts of spectrum and bandwidth resources that these massive numbers of IoT devices need. We envision that dynamic spectrum access (DSA) is to be needed now more than ever, as it is commonly viewed as a potential solution for overcoming such resource demand challenge.
Second, large numbers of newly emerging IoT applications are desperately in need for cloud offloading services due to their limited computation and storage capabilities, as well as to their low latency requirements. Empowering such IoTs with cloud offloading capabilities is therefore crucial to the successful support of key time-critical IoT applications like virtual reality, video surveillance, and precision healthcare, just to name a few.
Third, current cellular systems are designed for users' profiles that are different from the services requested by IoTs. In fact, current mobile systems are designed for limited numbers of connections and high-rate downlink data traffic, whereas IoTs require massive numbers of connections mostly for low-rate uplink traffic with various delay constraints~\cite{mehaseb2016classification}.~\\
\indent In this paper, we discuss some potential solutions that can be used to overcome these aforementioned challenges. Specifically, we leverage three key technology enablers, D2D, compressive sensing, and edge cloud, to address bandwidth resource shortages and network edge traffic bottlenecks that 5G systems face.
The potential of some of these technologies has already been recognized by ongoing research efforts such as the METIS project led by different research groups from various telecom companies. The main contribution of this work lies in the exploitation of key sparsity properties that are inherent to dynamic spectrum access and IoT traffic to develop efficient techniques that offer better IoT connectivity, alleviate congestion bottlenecks at network edges, and enable efficient dynamic wideband spectrum access and sharing.
The paper also identifies some open research challenges that still need to be overcome in order to enhance 5G's support of IoTs.

We want to mention that even though this work focuses on the support of IoTs via cellular/5G systems, depending on the IoT application, IoT devices can also be connected via various other means, such as WiFi, ZigBee, LoRa, etc~\cite{palattella2016internet}. In fact, recent studies~\cite{cisco2017forecast} show that by 2021, only about 7\% of IoT devices will be connected via cellular systems.

The remainder of this article is organized as follows. Section~\ref{sec:challenges} discusses IoT connectivity challenges. Section~\ref{sec:sparsity} presents the different approaches that can be used to exploit 5G sparsity to overcome these challenges. Finally, Section~\ref{sec:open} presents some open challenges and new research directions.

\section{Challenges 5G Faces in Support of IoTs} \label{sec:challenges}
\begin{figure}[!ht]
\centering{
\includegraphics[width=1\columnwidth]{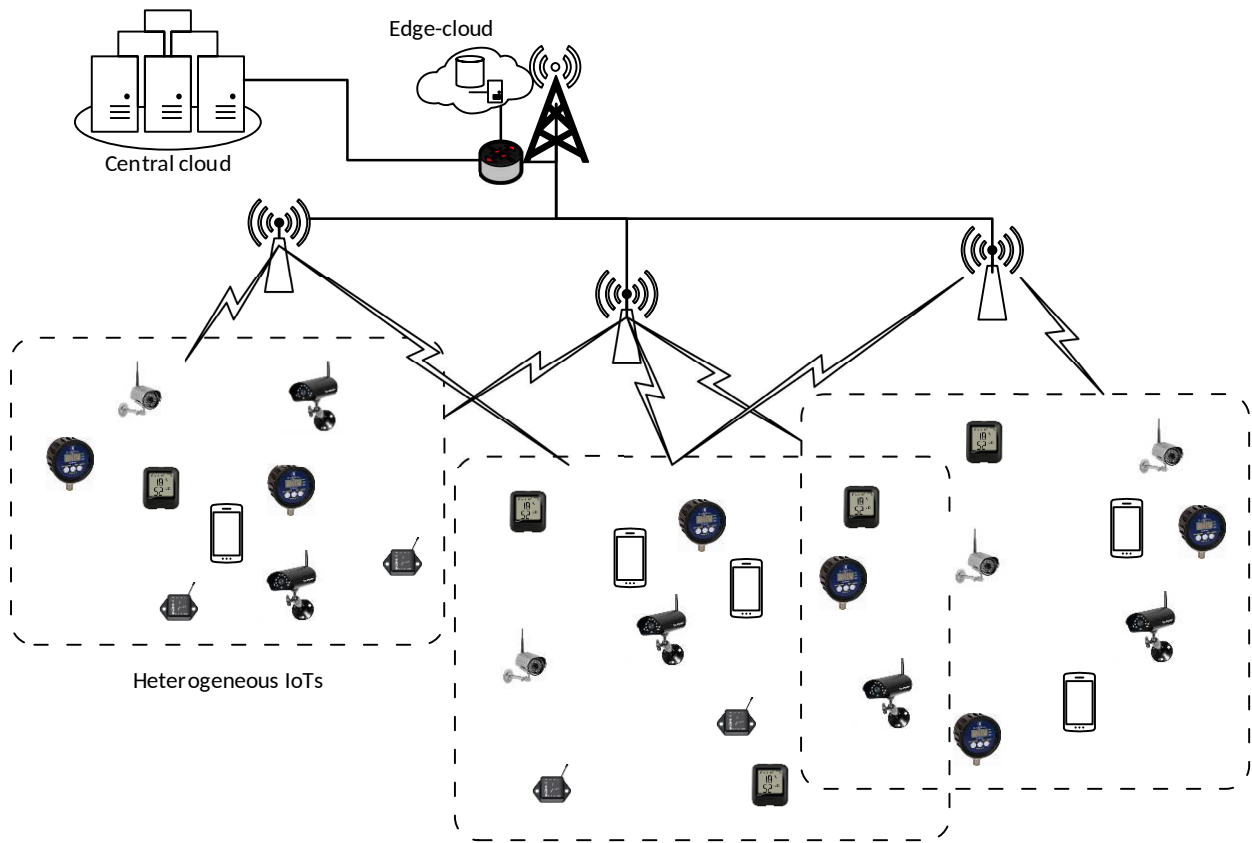}
\caption{5G support to massive and heterogenous IoTs with different service requirements enabled via edge-cloud or central cloud.}
\label{fig:sys}}
\end{figure}

Besides accommodating enhanced broadband mobile communications, 5G is anticipated to support a wide range of IoT applications with various heterogeneous requirements~\cite{3GPP-2016}. Fig.~\ref{fig:sys} illustrates 5G support to diverse IoT devices, where base station is augmented with edge cloud services.
The traffic generated by such IoTs is different from that generated by cellular systems in many aspects. First, unlike the case of broadband access, most of the IoT traffic is in the uplink. Moreover, IoTs' messages are typically small in size and sparse in time.
Furthermore, IoT devices are limited in energy and computation resources. These IoT devices' characteristics make their access to 5G systems different from classical cellular devices.
Given these traffic characteristics and resource constraints, IoTs can be classified, as illustrated by Table~\ref{tab:1}, into three classes based on their required services.

\begin{itemize}
  \item { \bf Massive IoTs (mIoTs)}: This class of IoTs includes large numbers of low-power, low-cost devices that generate low-rate, small-sized, delay-tolerant uplink traffic.
       Examples of such mIoTs are those envisioned for smart cities, smart homes, smart parking, environmental monitoring, etc.

  \item { \bf Ultra-reliable, low-latency IoTs (uIoTs)}: These IoTs need very low latency, high availability, and high reliability, but do not require high data rates. Examples of uIoTs are those envisioned for vehicle-to-everything (V2E) services, emergency management, remote healthcare, manufacturing control, smart grids, etc.

 \item{ \bf Hybrid IoTs}: These IoTs require both high data rates and low latency and are used in applications like virtual/augmented reality, video surveillance, law enforcement, etc.
 \end{itemize}

The connectivity of these heterogeneous IoTs will be ensured through heterogeneous networks with diverse ranges and data rates including cellular systems, WiFi, Bluetooth, Zigbee, Z-Wave, Sigfox, LoRA, Weightless, etc~\cite{palattella2016internet}. In particular, we focus in this work on cellular systems as there have been significant efforts towards developing standards relative to IoTs such as LTE-MTC, NB-LTE-M, and NB-IoT. With that being said, it is worth acknowledging that it is anticipated that only a small portion (about 7\%) of IoTs will be connected through cellular systems by 2021~\cite{cisco2017forecast}.

\subsection{Wideband Dynamic Spectrum Access Challenges}\label{subsec:21}
Serving these diverse, massive and heterogeneous IoTs calls for the development of new intelligent resource management approaches. Of particular importance is the need for efficient spectrum resource usage and access at higher frequency ranges.
Spectrum regulation agencies such as FCC and Ofcom have already issued notices for considering and using millimeter wave spectrum with the aim to meet the enormous bandwidth needs these massive IoTs are anticipated to require. Though spectrum policy makers are taking the necessary steps towards enabling and opening up wideband spectrum for 5G access, much remains to be done when it comes to developing resource allocation techniques. Although there is a general consensus among the researchers that dynamic spectrum access (DSA) will be a key for enabling efficient spectrum resource sharing at the millimeter wave range, there are key challenges that need to be addressed to be able to enable wideband DSA. Wideband spectrum sensing is one of such challenges that we focus on in this paper.

Spectrum sensing is the process by which unlicensed spectrum users identify unused portions of the licensed spectrum to use opportunistically. Despite the huge research efforts dedicated to developing efficient sensing techniques, not much has been done when it comes to exploiting the sparsity properties that are intrinsic to the wideband spectrum access. As will be discussed later, intelligently extracting and exploiting the sparsity properties inherent to the heterogenous occupancy nature of the wideband spectrum can significantly improve the sensing efficiency of the available portions of the wideband spectrum, and thus increase the overall spectrum utilization.

\begin{figure}[!ht]
\centering{
\includegraphics[width=1\columnwidth]{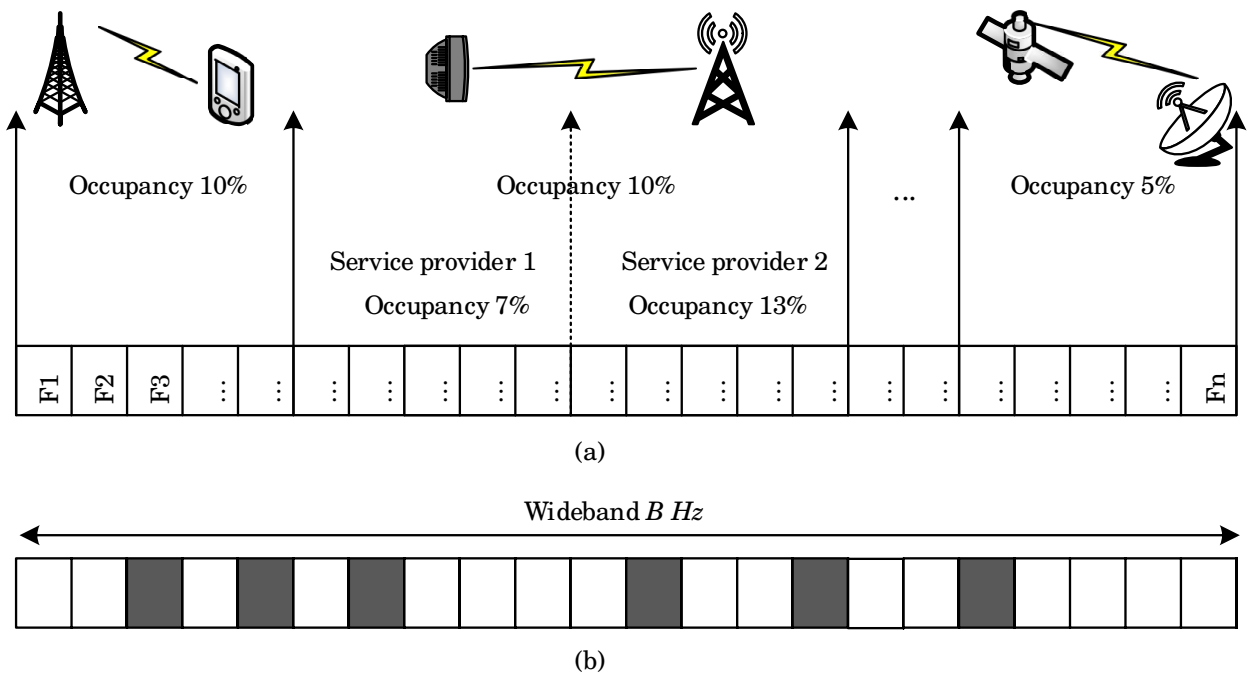}
\caption{$n$ frequency bands occupied by heterogeneous applications with different occupancy rates. The grey bands are occupied by primary users while the white bands are vacant. (a) is the statistical allocation while (b) is a realization of allocation in a given region at a given time slot. (a) reveals that the wideband spectrum is stochastically under-utilized and (b) is an instantaneous realization of this under-utilization.}
\label{fig:band_ocup}}
\end{figure}

With the opening up of the wideband spectrum access recently enabled by spectrum policy makers, and with  the device characteristics and traffic heterogeneity nature of these massive IoTs, traditional single-band spectrum sensing approaches are no longer effective and hence, new sensing approaches need to be developed.
The main reason for why traditional approaches are not efficient for wideband DSA is that they do require high numbers of sensing measurements; that is, in order to fully recover spectrum occupancy information, high (Nyquist) sampling rates are required, which can incur significant sensing overhead in terms of energy, computation, and communication.
Motivated by the sparsity nature of spectrum occupancy and in an effort to address the overhead caused by these high sampling rates, researchers have recently been focusing on exploiting compressed sampling theory to develop wideband spectrum sensing approaches that can recover information at sub-Nyquist sampling rates~\cite{chen2014survey}.
In Section~\ref{subsec:3a}, we present a novel wideband spectrum sensing technique that extracts key sparsity properties inherent to the wideband spectrum occupancy heterogeneity nature~\cite{yilmaz2016determination}  and exploits them through compressive sensing theory to improve the efficiency of spectrum sensing. An illustration of the wideband spectrum occupancy is shown in Fig.~\ref{fig:band_ocup}.

\subsection{Network Edge Traffic Challenges}\label{subsec:energy}
Cloud offloading has been adopted as a potential solution for overcoming the resource limitation of IoT devices, as it exempts them from having to deal with the computation, storage, and device-to-network communication burdens resulting from the running of the IoT applications.
Researchers have recently started exploring new ways to take cloud offloading to a higher level: bring cloud computing infrastructures closer to end-users, leading to what is commonly known today as Edge Clouds or Cloudlets.
Enabling IoT devices with edge cloud offloading capabilities is a key requirement for the 5G network architecture, crucial to successfully supporting IoT applications at scale, characterized with diverse and more stringent QoS requirements.
In addition to relieving the device from having to run its application locally, edge cloud offloading eliminates the need for having to send massive amounts of IoT data through the Internet, thereby generating lesser Internet congestions and, more importantly, improving IoT device responsiveness, essential to the support of time-critical IoT applications, such as realtime video surveillance, augmented reality, and remote health care.

With edge cloud offloading, IoT devices can replicate their memory objects (often small-sized) and transfer them to their associated Cloudlets. Despite these apparent resource elasticity benefits of edge cloud offloading, the massive numbers of devices each 5G cell is expected to support will render the network edges of 5G major traffic bottlenecks, thereby significantly limiting cloud offloading performance gains~\cite{biral2015challenges}.
In Section~\ref{subsec:3b}, we present techniques that leverage existing technologies such as D2D and compressive sensing theory to exploit key sparsity properties unique to IoT to alleviate congestion bottlenecks and overcome access scalability issues at 5G network edges.

\section{Extracting and Exploiting Sparsity for Efficient IoT Support in 5G} \label{sec:sparsity}
Having identified some key challenges facing the adoption of IoTs in 5G, we now present potential solutions that leverage compressive sensing theory to overcome these challenges.

\subsection{Enabling Efficient Wideband Spectrum Sensing}\label{subsec:3a}
In order to serve the massive numbers of IoTs, spectrum sensing techniques suitable for wideband spectrum access and sharing need to be carefully developed. We discussed in Section~\ref{subsec:21} the shortcomings of conventional sensing approaches when applied to wideband spectrum sensing. More specifically, the key limitations of such existing approaches lie in their high sampling rates and hardware capabilities needed to be able to recover sensing information for wideband spectrum access.
However, since (wideband) spectrum is heavily under-utilized~\cite{yilmaz2016determination} in that the number of occupied bands is significantly less than the total number of bands (i.e., the vector representing spectrum occupancy information is sparse), compressive sensing theory is an ideal candidate for fully recovering spectrum occupancy information while using sampling rates lower than sub-Nyquist rates~\cite{qin2015wideband}.
In other words, the recovery of the (sparse) spectrum occupancy vector can be done with a fewer number of sensing measurements.

With compressive sensing, the occupancy information of a spectrum consisting of $n$ bands can be recovered with only $m=O \big(k\log(n/k)\big)$ measurements where $m<n$ and $k$ is the number of occupied bands, referred to as the sparsity level.
%
The spectrum occupancy information vector, $\boldsymbol{x}_{n\times 1}$, is then recovered by minimizing the $\ell_0-$norm of $\boldsymbol{x}_{n\times 1}$ subject to a constraint on the error $\|\boldsymbol{y}_{m\times 1}-\Phi \Psi\boldsymbol{x}_{n\times 1}\|_{\ell_2}^2$, where $\boldsymbol{y}_{m\times 1}$ is the vector representing the $m$ measurements, $\Phi$ is a full-rank sensing matrix, and $\Psi$ is the discrete inverse Fourier transform Matrix. Due to its NP-hardness nature, recovery heuristics (e.g., $\ell_1$-norm minimization and orthogonal matching pursuit~\cite{qin2015wideband}) have been proposed in the literature for solving such problems. From a practical viewpoint, the implementation of wideband spectrum sensing requires the use of $m$ amplifiers and then mixing the received amplified signals with pseudo-random waveforms at Nyquist rates. After that, an integrator is applied followed by an analog-to-digital converter that takes samples
at sub-Nyquist rate. This architecture is known as analog-to-information converter (AIC) sampler~\cite{qin2015wideband}.

An observation we make by investigating the existing compressive sensing-based approaches is that they consider that the occupancy of wideband spectrum is {\em homogenous}, meaning that the entire wideband spectrum is considered as one single block with multiple bands, and the sparsity level is estimated across all bands and considered to be the same for the entire wideband spectrum.
However, in wideband spectrum assignment, applications of similar types (TV, satellite, cellular, etc.) are often assigned bands within the same block, suggesting that wideband spectrum is {\em heterogeneous}. That is, band occupancy patterns are not the same across the different blocks, since different application/user types within each block can exhibit different traffic behaviors, and hence, wideband spectrum occupancy may vary significantly from one block to another as illustrated in Fig.~\ref{fig:band_ocup}.
This trend has indeed also been confirmed by recent measurement studies~\cite{yilmaz2016determination}.

\begin{figure}[!ht]
\centering{
\includegraphics[width=1\columnwidth]{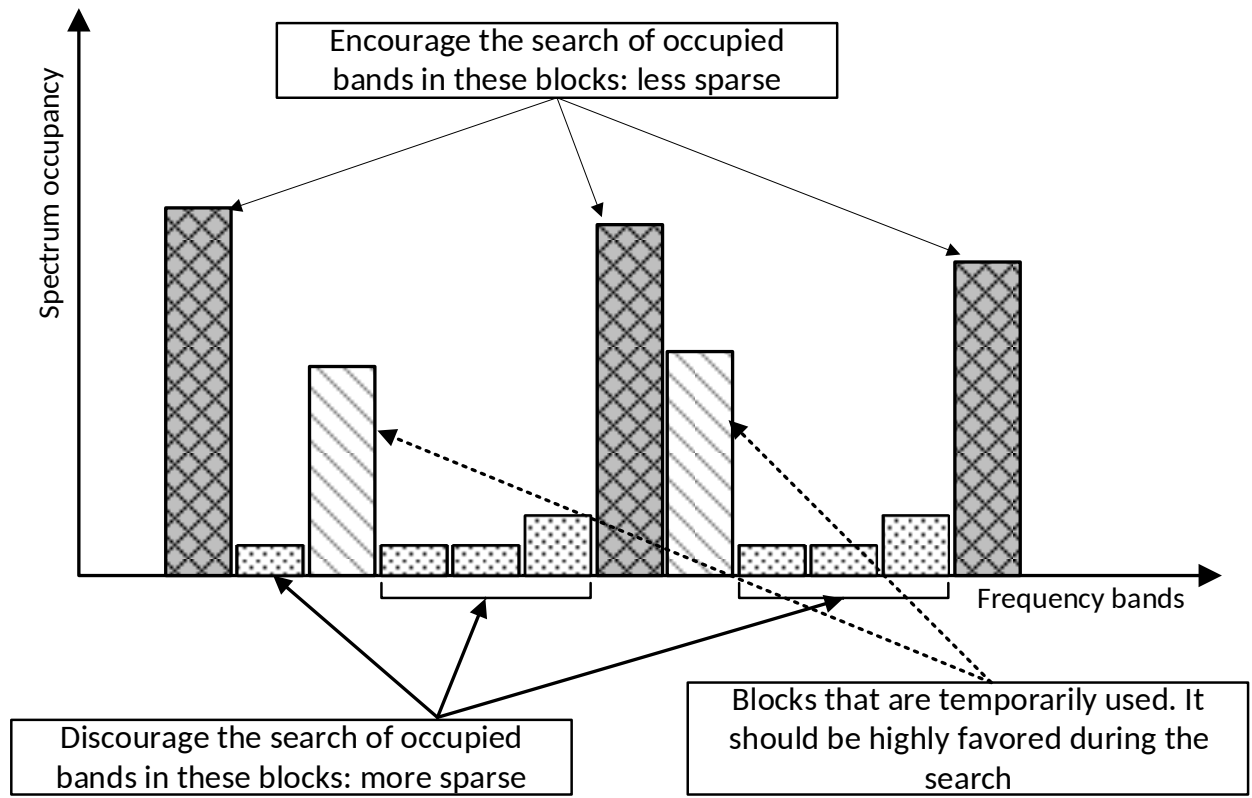}
\caption{Sparsity-promoting wideband spectrum sensing.}
\label{fig:band_ocup1}}
\end{figure}

\begin{figure}[!ht]
\centering{
\includegraphics[width=1\columnwidth]{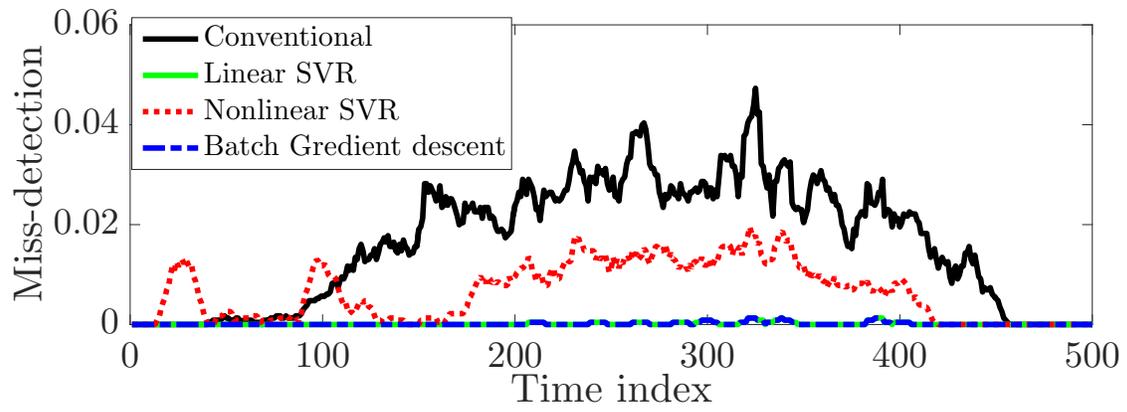}
\caption{Miss-detection performance evaluation of weighted compressive sensing under different regression techniques compared to conventional approach~\cite{qin2015wideband}. }
\label{fig:MD}}
\end{figure}

Incorporating this fine-grained sparsity structure into the formulation of wideband spectrum occupancy recovery allows us to improve the recovery performance and enhance the detection accuracy of wideband spectrum sensing. Specifically, such a block-like sparsity structure allowed us to formulate the problem as a weighted $\ell_1-$minimization problem, thereby resulting in an algorithm that provides faster spectrum occupancy recovery with lesser sensing overhead~\cite{khalfi2017exploiting}. 
The basic idea behind our algorithm is that the spectrum blocks that are more likely to be occupied are favored during the search. In addition, blocks corresponding to critical applications or for which some occupancy information is known are captured through careful design of block weights~\cite{khalfi2017exploiting}.
In essence, any additional knowledge about spectrum utilization behavior can be incorporated and exploited so that faster recovery of spectrum occupancy information can be achieved. Fig.~\ref{fig:band_ocup1} illustrates some of these design elements, where the blocks that are more likely to be unoccupied are encouraged to be sparser than the blocks which are more likely to be occupied.
Since the number of occupied bands changes over time, the design of the weights $\boldsymbol{w}$ can be based on the average occupancy of every spectrum block $i$, $\bar{k}_i$, such that $w_i=1/\bar{k}_i$~\cite{khalfi2017exploiting}. Furthermore, if each block occupancy, $\hat{k}_i$, can be estimated through learning (e.g., using regression techniques), better performance can be achieved when setting $w_i=1/\hat{k}_i$.
In Fig.~\ref{fig:MD}, we show the performance of the proposed weighted compressive spectrum sensing approach with band occupancy prediction (using different regression models), and compare it to a conventional wideband spectrum sensing approach~\cite{qin2015wideband}. Note that in the non-cooperative case where IoTs perform wideband spectrum sensing individually, there is no signaling overhead (information exchange with the other network entities to perform this task). However, in the cooperative case where multiple IoTs are involved in the sensing task, the signaling overhead becomes proportional to the number of cooperating IoTs. Table~\ref{tab:2} shows the signaling overhead associated with each of the approaches discussed in this paper.

\subsection{Overcoming Network Edge Traffic Bottlenecks}\label{subsec:3b}

As discussed in Section~\ref{subsec:energy}, the massive IoT traffic that 5G cells are required to support to enable edge cloud offloading will create severe congestion bottlenecks at the 5G network edges. One possible solution proposed in~\cite{abdelwahab2016replisom} to overcome this challenge lies in leveraging D2D and compressive sensing theory to reduce the number of connections established between the base stations and the IoTs, and to reduce the amount of offloading traffic. D2D communication technology has been adopted in LTE-advanced systems but only for public safety communications. When appropriately exploited, D2D can offer great advantages. Higher throughput, low latency, better availability and new services among other advantages make D2D an ideal candidate to help in the adoption and support of IoTs by 5G.

With the use of compressive sensing, instead of having all IoTs push their data to the base stations, the base stations can pull the data from only a subset of devices and use compressive sensing to recover the data of all IoTs. Here the sparsity that allowed the exploitation of compressive sensing comes from the fact/assumption that at a given time, only a few IoT devices experience changes in their memory and hence only few will need to upload their memory updates to the edge clouds.
Specifically, considering mIoTs with delay-tolerant requirements, every node multicasts to its neighbors a weighted value of the updated data replica with a defined coefficient that corresponds to the coefficient of the sensing matrix. When a node receives the weighted data replicas from other nodes, it adds its corresponding update, if any, and multicasts it during its time slot. After exchanging the data replicas, the nodes turn to the sleep mode for energy saving purposes. The base station pulls the measurements from few nodes, compared to the total number of mIoTs, where the number of these nodes should satisfy a condition that depends on the total number of nodes and the number of nodes that have data updates. Since most of the mIoTs have no update, then the vector corresponding to the memory replicas is sparse. Compressive sensing theory can accurately recover the data replicas for each IoT and support the corresponding ones through cloud services. The proposed protocol shows that signaling overhead is considerably reduced (only $m$ connections are established with BS), congestion is avoided and latency is improved by placing cloud services at the edge. The main shortcoming of this approach is that it only works with homogeneous IoTs and assumes a fixed sparsity level.

Potential improvements can be achieved through weighted compressive sensing as discussed previously in the wideband spectrum sensing context. In addition, learning and prediction approaches can also be used in conjunction with the recovery approach to improve the performance. This has been considered in~\cite{wang2012data} where a data gathering approach is proposed based on compressive sensing. The proposed scheme takes advantage of the correlation between data and introduces an autoregressive (AR) model in the recovery approach. IoTs can also be leveraged for performing wideband spectrum sensing but with a focus on reducing the reported measurements' cost~\cite{gwon2013scaling}, which can be combined with the work of~\cite{wang2012data}. Under the assumption that the sensed signal is sparse, a sparser basis can be found and can lead to a more compressed signal than the frequency domain basis. The IoTs report the measurements to network nodes that perform simple addition of the measurements coming from the IoTs and report them to the base station. This way, a constant communication cost is maintained (communication overhead is proportional to the number of network nodes). At the base station, the different measurements are exploited to recover the wideband spectrum occupancy.

Table~\ref{tab:2} summarizes the main proposed works that exploit sparsity features to enable the support of massive and heterogeneous IoTs.

\section{Open Research Problems and Directions}
\label{sec:open}
Despite the efforts made in exploiting the hidden sparsity structure for supporting IoTs through cellular systems, there remains key challenges that need to be overcome. We summarize here some of the research directions that we believe are worth investigating in the future.

{\bf Wideband spectrum occupancy behaviors}: Although some research efforts aiming to exploit spectrum occupancy sparsity to reduce traffic overhead have already been made, these approaches are either generic (not specific to wideband spectrum access) or achieve limited performance improvements due to the assumptions made. For instance, the spectrum occupancy heterogeneity structure inherent to dynamic wideband spectrum access is a feature that when exploited properly can allow for the design of more efficient compressive sampling approaches~\cite{khalfi2017exploiting}.
Also, a common limitation of these existing approaches lies in the fact that the spectrum occupancy sparsity level is considered constant and does not change over time.
Therefore, designing efficient recovery algorithms that exploit such features and structures in spectrum occupancy, finding bounds on the minimum required number of sensed measurements, and deriving error bounds on the achieved performances are some important challenges that remain open and hence require further investigation.

{\bf Cooperative wideband spectrum sensing}: As the demand for spectrum resources continues to rise with the emergence of 5G, devising efficient techniques for enabling dynamic spectrum sharing and access of wideband spectrum resources is needed more than ever. Of particular importance is cooperative spectrum sensing.
Considering and studying cooperative wideband spectrum sensing approaches under the observed heterogeneous structure of the spectrum occupation has great potential for improving spectrum sensing accuracy and reducing sensing overhead and is still an open research problem that requires further investigation.
Although this problem can be casted as a low-rank matrix minimization, deriving theoretical performance that consider tradeoffs between sensing overhead (delay, energy, etc.) and sensing accuracy while accounting for the time-variability of the spectrum occupation has not been investigated.

{\bf IoT heterogeneity}: Although there have been some research efforts that aim to leverage compressive sampling to reduce traffic jams, more work remains to be done when it comes to incorporating QoS.
The heterogeneity nature of IoT devices and their applications mandate that different IoT types may come with different QoS requirements. For instance, what IoTs designated for smart vehicle applications need is different from what those designed for collecting agriculture data (e.g., air temperature, humidity, soil moisture, etc.) need. QoS with respect to IoT is an area that has not received much attention, and hence, there remains challenges that need to be overcome.

{\bf Energy harvesting}: Energy availability and consumption continue to present a major challenge for IoTs due to their limited energy resources. Relying on energy harvesting and dedicated wireless energy transfer technologies emerge as key solutions to such challenges. Although there have recently been a research focus on developing energy harvesting techniques for wireless systems in general, not much has been done when it comes to developing energy harvesting techniques aimed for IoT devices and applications.

{\bf Security and privacy}: Most of these compressive sampling-based data reporting techniques that have been  proposed so far do not address security and privacy concerns. For instance, users' privacy may not be protected in the data reporting process. Existing traditional encryption protocols cannot be applied as they are to these proposed approaches, and hence, new privacy-preserving techniques need to be carefully designed so that compressive sampling can be exploited to reduce traffic, yet without compromising the privacy of users involved in the data reporting process.

\section{Conclusion}
IoTs have recently gained tremendous research attention as they are the main driver for a wide range of various applications. Of particular interest is the focus on leveraging compressive sampling theory and D2D communication technology to exploit some sparsity structures inherent to the spectrum resource access and sharing in 5G to overcome key challenges that 5G systems will face. Specifically, we focus on two key challenges that pertain to the support of IoTs in 5G: spectrum resource availability and traffic jams at network edges. We present some potential solutions for overcoming these two challenges, and identify some open research problems that remain to be addressed.

\section{Acknowledgement}
This work was supported in part by the US National Science Foundation (NSF) under NSF award CNS-1162296.

\bibliographystyle{IEEEtran}
\bibliography{References}

\begin{IEEEbiographynophoto}{Bassem Khalfi (S'14)} %
is currently a Ph.D student at Oregon State University.
His research focuses on various topics in the area of wireless communication and networks, including dynamic spectrum access, energy harvesting, and IoT.
\end{IEEEbiographynophoto}
\begin{IEEEbiographynophoto}
{Bechir Hamdaoui (S'02-M'05-SM'12)}
is an Associate Professor in the School of EECS at Oregon State University.
His research interest spans various areas in the fields of computer networking, wireless communications, and mobile computing.
\end{IEEEbiographynophoto}
\begin{IEEEbiographynophoto}
{Mohsen Guizani (S'85-M'89-SM'99-F'09)}
is currently a Professor in the ECE Department  at the University of Idaho, USA.
His research interests include wireless communications and mobile computing, computer networks, mobile cloud computing, security, and smart grid.
\end{IEEEbiographynophoto}

\newpage

\newpage

\begin{table}[!ht]
  \centering
    \caption{Classification of heterogeneous IoTs}\label{tab:1}
 \begin{tabular}{|c|c|}
    \hline
    Type/class of IoTs& Service characteristics\\ \hline \hline
    Massive IoTs (mIoTs) & Requires scalable connectivity and generates\\
     & small-sized, delay-tolerant uplink traffic\\ \hline
    ultra-IoTs (uIoTs)   & Requires reliable, low-latency, and highly\\
    & available network  connections  \\ \hline
    Hybrid & Includes low-latency, high-rate services\\
    \hline
  \end{tabular}
\end{table}

\newpage

\newpage

\newpage

\newpage

\begin{table*}[!ht]
  \centering
  \caption{Summary of the techniques exploiting sparsity.}\label{tab:2}
  \begin{tabular}{|c|c|c|c|c|}
    \hline
    \textbf{Ref.}& \textbf{Application}& \textbf{Comments} & \textbf{Sparsity}&\textbf{Signaling overhead} \\  \hline  \hline
     \cite{qin2015wideband} & Wideband spectrum sensing & without cooperation & fixed sparsity &no signaling overhead\\  \cline{3-3} \cline{5-5}
     &  & with cooperation & & proportional to number of SUs\\ \hline
     \cite{khalfi2017exploiting}& Wideband spectrum sensing& Exploited spectrum heterogeneity & varying sparsity &no signaling overhead\\ \hline
     \cite{abdelwahab2016replisom} & Upload memory replicas of mIoTs & Exploited D2D communications & fixed sparsity &$m$ connections with the BS\\ \hline
    \cite{wang2012data} & Adaptive data gathering & Exploited correlation & varying sparsity &$m$ transmissions to the sink\\ \hline
    \cite{gwon2013scaling}& Compressed measurement reporting & Exploited signal's stationarity & fixed sparsity &for $p$ active IoTs and $n$ network\\
    & & + D2D & &nodes: $n+p$ transmissions\\ \hline
    \cite{wang2012sparsity} & Wideband spectrum sensing & Two-step approach to adjust & varying sparsity &no signaling overhead\\
    & & the number of measurements & & \\ \hline
    \end{tabular}
\end{table*}

\end{document}